\documentclass[prx,twocolumn,longbibliography,superscriptaddress,preprintnumbers]{revtex4-2}
\usepackage[colorlinks,bookmarks=true,citecolor=blue,linkcolor=blue,urlcolor=blue]{hyperref}
\usepackage{dcolumn,graphicx,amsfonts,amsthm,bm,color,appendix,float,tabularx}
\usepackage{braket}
\usepackage[normalem]{ulem}
\usepackage[version=4]{mhchem}
\usepackage{siunitx}
\usepackage{CJKutf8}
\usepackage{multirow}
\usepackage{booktabs}
\usepackage{xcolor}
\usepackage{soul}
\usepackage{graphicx}
\usepackage{braket}
\usepackage[capitalize]{cleveref} 
\usepackage{changes}
\definechangesauthor[name=Antoine, color=blue]{ag}

\newcommand{\NNG}[1]{{$G$Net#1}}

\newcommand{\CCQ}{Center for Computational Quantum Physics, Flatiron Institute, 162 5th Avenue, New York, NY 10010, USA}

\begin{document}

\title{Neural-Network Quantum Embedding Solvers for Correlated Materials}

\author{A. Valenti}
\affiliation{\CCQ}

\author{I. Park}
\affiliation{\CCQ}

\author{A. Georges}
\affiliation{\CCQ}
\affiliation{Coll\`{e}ge de France, 11 place Marcelin Berthelot, 75005 Paris, France}
\affiliation{CPHT, CNRS, Ecole Polytechnique, IP Paris, F-91128 Palaiseau, France}
\affiliation{DQMP, Universit\'{e} de Gen\`{e}ve, 24 quai Ernest Ansermet, CH-1211 Gen\`{e}ve, Suisse}

\author{A.J. Millis}
\affiliation{\CCQ}
\affiliation{Department of Physics, Columbia University, 538 West 120th Street, New York, NY 10027, USA}

\author{O. Parcollet}
\affiliation{\CCQ}
\affiliation{Universit\'e Paris-Saclay, CNRS, CEA, Institut de Physique Th\'eorique, 91191, Gif-sur-Yvette, France}

\date{\today}

\begin{abstract} 
Quantum impurity solvers are the computational bottleneck of quantum embedding
approaches to correlated materials, such as dynamical mean-field theory (DMFT).
We show that neural networks trained on synthetic, material-agnostic data learn
the impurity mapping from hybridization functions and local interactions to
Green’s functions with quantitative accuracy for both model systems and
real materials, providing fast solvers for single- and multi-orbital models. Benchmarks against
numerically controlled quantum Monte Carlo show that the method
reproduces the Mott transition, multi-orbital phase diagrams of
Hubbard-Kanamori models, and the electronic properties of SrVO$_3$ and
SrMnO$_3$. The learned solvers achieve orders-of-magnitude speedup and can
initialize controlled calculations, dramatically accelerating DMFT while
preserving accuracy.
\end{abstract}

\maketitle

The {\it many electron problem} \cite{dirac1929}, one of the grand challenges of
modern science, consists of computing experimentally
measurable correlation functions of a system of interacting electrons in a solid,
given some basic knowledge of atomic structure and some external parameters such as chemical potential and temperature.
Due to the exponential growth of Hilbert space with system size, it is a notoriously hard high-dimensional problem
whose solution has motivated decades of development of theoretical and computational methods. 
Neural networks (NN) offer a promising approach to this challenge
as their architectures efficiently approximate high-dimensional functions,
capturing complex correlations and exhibiting strong generalization from data.

Quantum embedding methods such as Dynamical Mean Field Theory (DMFT)~\cite{georges1992hubbard,georgesRMP1996,kotliarRMP2006} form a
well-established theoretical framework for studying the electronic structure of
strongly correlated materials and related theoretical models. They reduce the complex
many-electron problem to solving  an auxiliary `quantum impurity' (QI)
model, consisting of a finite number of localized orbitals or sites coupled to
a self-consistently determined, non-interacting electronic bath.
While QI `solvers' -- algorithms~\cite{gull2011continuous,werner2006continuous,
werner2006hybridization,rubtsov2005continuous, gull2008continuous,
bulla2008numerical, bulla1999zero, white1992density, schollwock2005density,
ostlund1995thermodynamic} for solving quantum impurity models-- are available,
they 
have a significant computational cost. Reducing this cost could transform the field by 
speeding up computations of strongly correlated materials 
with DMFT combined with density functional theory (DFT+DMFT) 
~\cite{anisimov1997first,
lichtenstein1998,kotliarRMP2006}
to the level of DFT computations routinely used for weakly correlated ones.

A quantum impurity solver is a high dimensional function, returning 
the impurity Green's functions from the parametrization of the bath and 
the values of the atomic interactions. 
In this letter, we show that neural networks trained on simulated ground-truth data 
provide a powerful, flexible and general representation of quantum impurity solvers that
can be successfully used to compute the electronic structure of materials with strong correlations within DFT+DMFT.
We go beyond previous machine learning approaches to DMFT~\cite{arsenault2014machine, sheridan2021data,
sturm2021predicting,agapov2024predicting, lee2025language, mitra2025deep, rao2026},
by constructing a set of NN quantum impurity solvers for one-, two-, and three-orbital models
using synthetic training data obtained from systematically sampling bath parametrizations and interactions, 
independently of the electronic structure details of specific materials.
We benchmark our NN solvers in various DMFT
computations: the first-order Mott transition in the single-band Hubbard model,
the phase diagram of a two-orbital Hubbard-Kanamori model, and DFT+DMFT 
computations of two materials (SrVO$_3$, a correlated metal, and SrMnO$_3$, a Mott insulator). In all cases, we find
excellent agreement with controlled solvers, demonstrating that NN-based solvers 
trained on synthetic data perform reliably on a broad domain of applications.

Neural network solvers are extremely fast, requiring seconds even for
multiorbital materials compared to hours or days for controlled numerical
solvers such as continuous time quantum Monte Carlo
(CT-QMC)~\cite{werner2006continuous, werner2006hybridization,
gull2011continuous} or tensor networks~\cite{schollwock2005density,
ostlund1995thermodynamic}, thereby enabling numerous applications.
However, a lack of accuracy control is  inherent in machine
learning methods.   We show that controlled
accuracy can be obtained by using NN solvers as accelerators,
performing a few refinement iterations with an accurate solver starting from a
converged solution obtained efficiently with the NN solver.


{\it Setup---} A QI problem is entirely specified by: (i) a hamiltonian $H_{\rm int}$ describing 
local many-body interactions among the $M$ local orbitals defining the 
embedded atomic shell - with $m=1,\cdots,2M$ labeling both spin and orbital indices; 
(ii) a matrix of local levels $\epsilon^{d}_{mm'}$ (in which the overall 
chemical potential $\mu$ can be included) 
and (iii) a set of functions $\Delta_{mm'}(\tau)$ 
(the dynamical mean-field, or hybridization function) which encode the transfer of electrons 
between the local atomic shell and the self-consistent bath. 
Here, $\tau$ is the imaginary time which runs from $\tau=0$ to $\tau=\beta\equiv 1/T$, the 
inverse temperature. 
The precise formulation of the imaginary-time effective action describing the QI model is given in Appendix~\ref{app:Hint}, 
together with the Hubbard and Hubbard-Kanamori forms of the interaction hamiltonian $H_{\rm int}$ considered in this 
work, parametrized by the screened Coulomb interaction $U$ and Hund's coupling $J$. 
A QI solver is a high-dimensional function: 
\begin{equation}
\left( H_{\mathrm{int}},\beta,\,\,\epsilon^d_{mm'},\Delta_{mm'}(\tau) \right) \rightarrow G_{mm'}(\tau)
\label{eq:map}
\end{equation}
that returns the impurity Green's function $G_{mm'}(\tau)$ 
from the local levels $\epsilon^d_{mm'}$, the dynamical mean-fields $\Delta_{mm'}(\tau)$, 
and the parameters $H_{int},\beta$. 

In order to facilitate the
network's task, we use a parsimonious representation of the functions $\Delta$ and $G$  on the 
imaginary axis, as vectors $\{\Delta(\tau_i)\}$, $\{G(\tau_i)\}$, where $\{\tau_i\}$ with $1\leq i \leq N_\tau$ is
a compact Discrete Lehman Representation (DLR) mesh~\cite{kaye2022discrete}.
This grid uses a minimal number of points $N_\tau \sim \log(\beta \omega_\text{max}) \log(1/\epsilon)$
which increases logarithmically as a function of $\beta$, the high-energy cutoff $\omega_\text{max}$ (the maximum support of the spectral functions)
and the precision $\epsilon$. In this work, $N_\tau < 10^2$.

As we want to train our NN solver for a range of temperatures, 
using a different grid for each $\beta$ would a priori require 
the use of NN architectures that can deal with input data of different sizes.
We circumvent this challenge 
by noting that, from the spectral representation,  
the function $G(\tau)$ can be viewed as a function of $\tau/\beta$ with a high energy cutoff $\beta\omega_\text{max}$ 
(which may depend on interaction).
To represent our functions for a range of temperatures $[0,\beta_M]$, 
we therefore use the {\it same} DLR grid generated for $\beta=\beta_M$, cutoff $\omega_{\rm max}$ 
and precision $\epsilon$ 
(see Appendix~\ref{app:dlr}). 


The crucial ingredient in building a neural-network QI solver is 
the training dataset. Our aim is to construct 
a well-defined, general-purpose QI solver
that is usable in any DMFT computation
within its specified range of parameters (e.g. temperatures, interactions)-
rather than a solver applicable only to a specific class of materials or band structures.
To avoid material-specific bias, we purposefully utilize a {\it
synthetic} dataset, which we create in three steps.
First, we randomly draw a set ${\cal D}_1$ of hybridization functions using random coefficients
in the DLR representation in terms of a sum of $N_\tau$ exponentials~\cite{kaye2022discrete}.
Second, we solve the QI model for all $\Delta_j \in {\cal D}_1$ to obtain $G_j$ using a numerically exact
continuous time Quantum Monte-Carlo (CTSEG~\cite{werner2006continuous, gull2011continuous} for 1 orbital, or CTHYB for 2 and 3 orbitals).
In ML parlance, $\left( H_{\mathrm{int}},\beta,\,\,\epsilon^d_{mm'},\Delta_{mm'}(\tau)\right)_j$ 
constitute the input features, and $G_j$ are the corresponding labels.
Third, we augment our dataset by applying $1-2$  DMFT iterations 
with a simple self-consistency condition $\Delta = G$,
and randomly varying $H_\text{int}$ (and, if included, $\beta$) between iterations.
This augmentation is crucial for the quality of the NN solver.
Remarkably, we show below that the solver generalizes well and performs effectively for DMFT computation of 
materials where the self-consistency condition is far more complex.

We train a set of feed-forward neural network QI solvers 
for $M$ orbital systems with $M = 1,2,3$. In the following, we refer to them as \NNG{}.
The total number of training examples is $1.6 \times 10^4$ (resp. $1.8 \times 10^4, 8.6\times10^4$) 
for $M=1$ (resp. $M=2,3$).
We also train a solver for the density $n$ for $3$-orbital systems, as it yields 
more precise results than deducing $n$ from \NNG{}.
Details of the neural networks, the training set, and the computational cost 
are presented in Appendix~\ref{app:NN}. 

\begin{figure}[tb] 
    \centering
    \includegraphics[width=1.\linewidth]{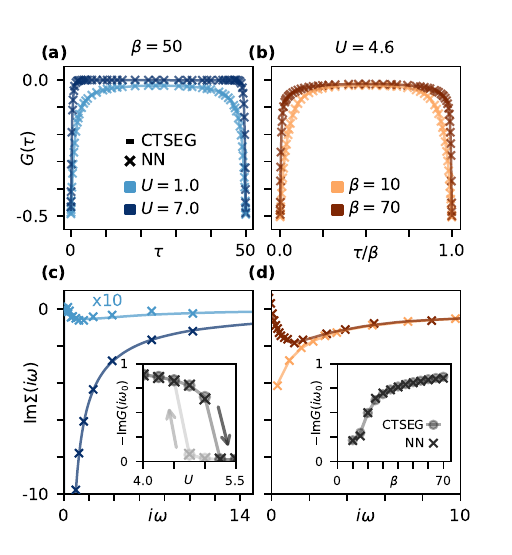}
    \caption{ 
    Comparison of the \NNG\ solver (symbols) with CTSEG (lines) for the half-filled one orbital DMFT solution of the Hubbard model on the Bethe lattice.
    Left column: $U$-scan at $\beta=50$. Right column: $\beta$-scan at $U = 4.6$.
    (a,c) $G(\tau)$ (a) and Im$\Sigma(i\omega)$ (c) for $U \in \{1,7\}$. On panel (c), for $U = 1$, Im$\Sigma(i\omega)$ is multiplied by 10 for readability. 
    (b,d) Same for $\beta = 10,70$. 
    Inset (c) $-\mathrm{Im}G(i\omega_0)$ vs $U$ showing DMFT metal-insulator transition hysteresis, with grey (black) arrows indicating insulating (metallic) branches.
    Inset (d) $-\mathrm{Im}G(i\omega_0)$ vs $\beta$, close to the Mott transition. 
    }
    \label{fig:1orb}
\end{figure}

{\it 1-orbital model---}
Fig.~\ref{fig:1orb} compares results obtained from our \NNG{} solver (trained at the range of $\beta\in[10,80]$)
to numerically exact results obtained with a CTSEG quantum Monte-Carlo algorithm~\cite{werner2006continuous, gull2011continuous},
for the conceptually simplest case: 
the DMFT non-magnetic solution of the half-filled one orbital Hubbard model 
on the Bethe lattice~\cite{georgesRMP1996} 
(semi-circular density of states with bandwidth normalized to $4$)
\footnote{Alternatively, this can be seen as the exact solution of a 
fully-connected Hubbard model with random hopping, in which case no magnetic phase is present\cite{georgesRMP1996}.}.
Additional results away from half-filling are presented in Appendix~\ref{app:1orb}. 
This model 
exhibits a paramagnetic metal to paramagnetic
insulator (``Mott") transition as the interaction strength $U$ is increased above a
critical value.  Panel (a) shows that the NN-computed imaginary time Green's
function is indistinguishable from the numerically exact ground truth data for
two very different interaction strengths; one below and one above the critical
value for the Mott transition. Panel (b) shows that the NN solver
 also correctly reproduces the temperature dependence even for an interaction
strength close to the Mott critical value~\cite{georgesRMP1996}. 
Figure~\ref{fig:1orb}(c,d) shows the corresponding self-energies,
obtained via the Dyson equation
\begin{equation*}
 \Sigma_{mm'}(i\omega_n)
= i\omega_n \delta_{mm'} - \epsilon^d_{mm'} - \Delta_{mm'}(i\omega_n) - G^{-1}_{mm'}(i\omega_n).
\end{equation*}
In both cases, we observe an excellent agreement between the two solvers.
The insets in Fig.~\ref{fig:1orb}(c,d) present a more detailed examination of the Fourier transform of $G$ evaluated at the lowest Matsubara frequency, $\omega_0=\pi T$. \NNG{} correctly reproduces the hysteresis (bistable behavior) occurring in the DMFT approximation to the interaction-driven Mott transition~\cite{georgesRMP1996} (inset in panel (c)) as well as the temperature dependent change from insulating (low $G$) behavior at high $T$ (small $\beta$) to metallic (high $G$) at low $T$ (inset in panel (d)).

{\it 2-orbital phase diagram---} 
We now show that \NNG{} reliably captures the richer physics of 
multi-orbital Hubbard-Kanamori models~\cite{kanamori1963electron, georges2013strong}.  
In the presence of crystal field splitting, the local interaction Hamiltonian
becomes \begin{align} H_{\rm int} = H_{\rm H-K}(U,J) + \sum_{\sigma}
\Delta_{\rm cf} (n_{1,\sigma}-n_{2,\sigma}), \end{align} 
where $H_{\mathrm{H-K}}$ is the Hubbard-Kanamori interaction Hamiltonian (see Appendix~\ref{app:Hint})
and $n_{i,\sigma}$ denotes the occupancy of orbital $i$ and spin $\sigma$. 
Two-orbital models arise for example in the context of materials involving
transition metals with partially filled $d$-shells. The key parameters in addition to
the density of states are a "charging energy" $U$, a multiplet-splitting energy $J$ (Hund's
coupling) and a crystal field splitting $\Delta_{\rm cf}$. At half-filling, the interplay 
between these energies leads to a sequence of orbital polarization
transitions~\cite{werner2007high}. 

\begin{figure}[tb]
    \centering
    \includegraphics[scale=1]{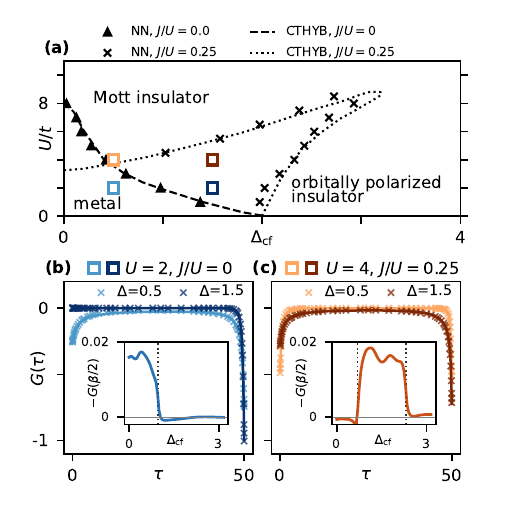}
    \caption{
    (a) Metal-insulator phase boundaries obtained using \NNG{} for 2-orbital at $\beta=50$ (markers), compared  with the CTHYB results from Ref.~\cite{werner2007high}. The colored squares denote the parameter values at which the Green's functions are directly compared with CTHYB in panels (b) and (c). Blue squares denote $J/U=0$, while orange/red squares denote $J/U=0.25$. Crosses in panels (b) and (c) denote the NN solutions, whereas solid lines denote the CTHYB results. The insets show the NN-predicted $G(\beta/2)$ along horizontal cuts in the phase diagram. The vertical dotted lines in the insets correspond to the position of the phase transition taken from Ref.~\cite{werner2007high}.
    }
    \label{fig:2orb}
\end{figure}

Using the \NNG{} solver for $M=2$ orbitals trained at $\beta=50$, 
we determine the DMFT phase diagram of this model on the Bethe lattice (defined as above).  
The metal-insulator transition is identified by
monitoring the value of $-G(\beta/2)$, and the results are compared with the CTHYB results of Ref.~\cite{werner2007high}, as shown in Fig.~\ref{fig:2orb}(a).
The insets in  Fig.~\ref{fig:2orb}(b,c) show this
criterion, $-G(\beta/2)$, at selected values of $U$, $J$ as a function of
$\Delta_{\rm cf}$. A jump indicates the position of the phase
transition. We find that the phase diagram is consistent with that of 
Ref.~\cite{werner2007high}, within a few percent
of critical parameter values.  
We further compare the Green's functions in panels (b) and (c) between the \NNG{} solver (crosses), fully converged using $\Delta$-mixing ($\Delta_{i+1} = \gamma \Delta_i + (1-\gamma)\Delta_{i-1}$, with $i$ the iteration number of the self-consistent loop), and the CTHYB results (solid lines), finding good agreement.

\begin{figure}[tb]
    \centering
    \includegraphics[width=0.5\textwidth]{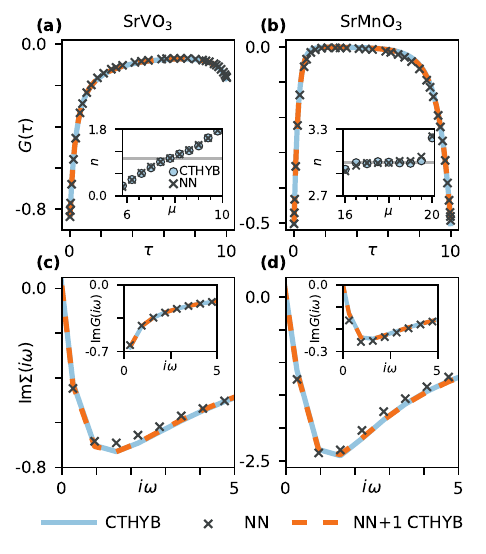}
 \caption{DFT+DMFT results for SrVO$_3$ (left) and SrMnO$_3$ (right) at $\beta = 10$ using
    \NNG{} (crosses) for $M=3$ orbitals at $\beta=10$, CTHYB (solid blue), and a converged computation with the
    \NNG{} solver followed by one iteration with CTHYB (NN+1$\times$CTHYB, dashed red). 
    For each material, we show $G(\tau)$ (top), Im$\Sigma(i\omega)$ (bottom), and
    Im$G(i\omega)$ (bottom, insets). 
    Top insets: impurity occupancy $n(\mu)$ with the NN density-solver (crosses) and CTHYB (circles).
    \label{fig:3orb}
 }
\end{figure}

{\it 3-orbital materials---}Finally, we demonstrate the application of \NNG{}
to DFT+DMFT calculations for real materials and how it accelerates
computationally demanding DFT+DMFT calculations. We show two real material
examples,  SrVO$_3$~\cite{fujimori1992evolution}, a moderately correlated
metal, and SrMnO$_3$~\cite{lee1995electronic, kim2010photoemission}, a strongly
correlated Mott insulator. The DMFT impurity model for both cases involves
three orbitals corresponding to the $t_{2g}$ multiplet of the V or Mn atoms
with cubic symmetry.
For these calculations, we use the \NNG{} solver for $M=3$ orbitals at
$\beta=10$ in combination with the NN density-solver (see Appendix~\ref{app:3orb}
for details.)

Figure~\ref{fig:3orb} compares the \NNG{} solver results (crosses) with a
CTHYB reference calculation (solid blue line). 
The agreement is excellent for the Green's function $G(\tau)$, while the self-energy
(obtained via the Dyson equation) exhibits good but slightly lower accuracy. These calculations
are performed at a fixed chemical potential $\mu$, which gives the target density determined from DFT calculation. The insets in the upper panels shows the impurity
occupancy $n(\mu)$ for various chemical potentials, computed with the NN density-solver 
at the final \NNG{} DMFT iteration. This solver predicts the density more accurately
than if directly computed from the \NNG{} result at the final DMFT iteration, 
and again shows very good agreement with CTHYB.
These results show that \NNG{}
provides accurate results for DFT+DMFT computations of two materials,
despite being trained only on 
hybridization functions without information on material specifications. 

Neural network QI solvers run very quickly (typically within one second), but
they lack any accuracy guarantee or error bars and, as seen for example in the
lower panels of Figure~\ref{fig:3orb} the results are not exactly identical to the
numerically exact results.
This issue motivates a {\it NN-accelerated
strategy}~\cite{arsenault2014machine,lee2025language}: use the NN solver to
rapidly obtain a converged solution, then refine and validate with a few
iterations of an exact but expensive solver.  Figure~\ref{fig:3orb} (dashed red
line) demonstrates this approach—a single CTHYB iteration initialized from the
\NNG{} solution yields self-energies in excellent agreement with the full
reference calculation (differences in $G$ and $\Sigma$ are shown in Appendix~\ref{app:3orb}).
%
Figure~\ref{fig:precision_svo} further illustrates the acceleration,
comparing the DMFT convergence for  three
physical quantities—impurity occupation $n_\text{imp}$,
self-energy $\Sigma$, and Green's function $G$ -- using two different initializations.
The NN-initialized calculation (crosses, red) converges substantially faster than
the conventional DMFT calculation (circles, blue), which starts from a DFT calculation and updates the chemical potential given the target density $n_{\mathrm{latt}}$ from the DFT. 
Since obtaining the initial NN solution adds negligible cost compared
to exact solver iterations, this strategy achieves an overall speedup
of approximately one order of magnitude in this example.              

\begin{figure}[tb]
    \centering
    \includegraphics[scale=1.0]{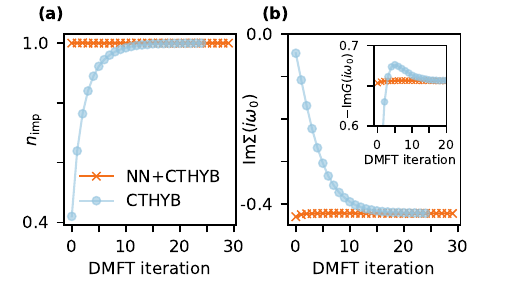}
    \caption{
  Convergence of 
   the impurity occupancy $n_{\mathrm{imp}}$ (a),
  $\mathrm{Im}\Sigma(i\omega_0)$ (b), 
  $\mathrm{Im}G(i\omega_0)$ (inset),
   vs the iterations of the DMFT self-consistency loop, using CTHYB,
   starting from a DFT solution (circle, blue) and updating $\mu$
   or a converged DMFT computation using \NNG{} and density solvers (cross, red), for the case of SrVO$_3$. Here,
   $i\omega_0$ denotes the first Matsubara frequency. 
}
    \label{fig:precision_svo}
\end{figure}

{\it Discussion---}
In this paper, we have presented neural-network-based solvers for an important class of quantum many-body systems:
the quantum impurity models central to quantum embedding approaches for the electronic structure of materials with
strong electronic correlations.
We have quantified their accuracy by benchmarking them against numerically exact quantum Monte Carlo results.

By constructing our training dataset independently of materials-specific
details — except for the form of the atomic interaction and the number of
orbitals — we obtain a generic quantum impurity solver directly
applicable to any DMFT computation within its specified parameter range.
The success of our solver in DFT+DMFT computations of SrVO$_3$ and SrMnO$_3$
indicates that our synthetic dataset is sufficiently rich to cover the
relevant hybridization function space encountered in realistic materials calculations.   

When used free-standing, the NN solvers achieve speedups of several orders of magnitude (seconds versus hours or days) 
compared to existing techniques while maintaining good accuracy.
When used to initialize a conventional calculation, they achieve acceleration by 1-2 orders of magnitude (1-2 iterations versus 10-30) 
with very high accuracy.
This neural-network-based acceleration has transformative potential for DFT+DMFT electronic structure calculations.

Future applications may include fast computation of total energies, forces, structural relaxation, phonon spectra,
large supercells with correlated atoms, and parameter space searches for materials design.
Various generalizations of neural network solvers should also be explored.  An important open question in this context is the generation of  synthetic data sets with appropriate coverage of the needed function space.
Near term research targets include extending the NN solvers to include spin-orbit coupling, more orbitals, and the spatial correlations incorporated in cluster dynamical mean field theory.
Also interesting are extensions to higher-order correlation functions such as response or vertex functions, real-time dynamics, and retarded interactions as in GW+DMFT. 

{\it Acknowledgements ---} The work of AJM at Columbia University was supported by Programmable Quantum Materials, an Energy Frontier Research Center funded by the U.S. Department of Energy (DOE), Office of Science, Basic Energy Sciences (BES), under award DESC0019443. 
We are grateful to Michel Ferrero, Rudy Morel, and Filippo Vicentini for useful discussions. IP and AV thank 
Harry LaBollita for technical support.
The Flatiron Institute is a division of the Simons Foundation.

{\it Data availability ---} Data and code are available upon reasonable request to the authors. 
A trained model for one orbital and an example calculation are publicly available~\cite{repo}.

%

\clearpage

\appendix
\twocolumngrid


\section{Interaction Hamiltonians \label{app:Hint}}

We consider the quantum impurity action given by 
\begin{multline}
S = 
-\int\!\!\!\! \int_0^\beta\!\! d\tau  d\tau'  \sum_{mm'} \sum_{\sigma}
d^\dagger_{m\sigma}(\tau) 
\Bigl[
(\partial_\tau - \epsilon_d )_{mm'} \delta(\tau-\tau') -
\\ 
\Delta_{mm'}(\tau-\tau') \Bigr] 
d_{m'\sigma}(\tau') 
+ \int_0^\beta\!\! d\tau\, H_{\rm int}[d^\dagger(\tau), d(\tau)]
\end{multline}
where $m, m'$ labels orbital and $\sigma$ labels spin index. 
$H_{\rm int}$ is the local interaction Hamiltonian of the Hubbard form for one orbital and Hubbard-Kanamori 
form with parameters $U$, $J$ for two and three orbitals. 
Here we restrict the formulation to a spin-diagonal and spin-independent (i.e. paramagnetic phase) hybridization function. 

For one-orbital systems, we use the Hubbard interaction Hamiltonian: 
 \begin{align}
 H_{\rm int} &= U{n}_{\uparrow}{n}_{\downarrow},
 \end{align}
 where $U$ is the on-site Coulomb interaction. 

For multi-orbital systems, we use the Hubbard-Kanamori interaction Hamiltonian:
\begin{align}
H_{\mathrm{H-K}} &= U\sum_{m}{n}_{m^{}\uparrow}{n}_{m^{}\downarrow}+U^{\prime}\sum_{m^{}\neq m^{\prime}} {n}_{m^{}\uparrow} n_{m^{\prime}\downarrow} \notag\\
&+(U^{\prime}-J)\sum_{m^{}<m^{\prime},\sigma = \uparrow, \downarrow}{n}_{m^{}\sigma}{n}_{m^{\prime}\sigma}
\notag\\
&- J\sum_{m^{}\neq m^{\prime}} d^{\dagger}_{m^{}\uparrow}d^{}_{m^{}\downarrow}{d}^{\dagger}_{m^{\prime}\downarrow}{d}^{}_{m^{\prime}\uparrow} \notag\\
&+J\sum_{m\neq m^{\prime}}{d}^{\dagger}_{m^{}\uparrow}{d}^{\dagger}_{m^{}\downarrow}{d}^{}_{m^{\prime}\downarrow}{d}^{}_{m^{\prime}\uparrow},
\label{eq:kanamori}
\end{align}
where $U$ is the intra-orbital Coulomb interaction, $U'$ the inter-orbital interaction, and $J$ the Hund's coupling. The last two terms of the interaction Hamiltonian corresponds to the spin-flip and pair-hopping term, both with strength $J$, and $U^{\prime} = U-2J$ is set from the rotational invariance of the local interaction. $m$ and $m^{\prime}$ are the orbital indices and $\sigma$ is the spin index. This Hubbard-Kanamori interaction Hamiltonian is used in both cases of 2-orbital and 3-orbital in our study.

\begin{table*}[tb]
\begin{ruledtabular}
\begin{tabular}{cccccc|cccc|c}
        \multicolumn{6}{c|}{Impurity model parameters} & \multicolumn{4}{c|}{Parameters for generating $\Delta$} & \\
\toprule 
$n_{\mathrm{orb}}$ &
$\beta$ & $U$ & $\mu_{\rm NN}$ & $J$ & $\Delta_{\rm cf}$ & $\epsilon$ & $\alpha_m$ & $u^k_{1,m}$ & $u^k_{2,m}$ & $N_{\rm train}$  \\
\colrule
 $1$ & $[10,80]$
& $[0,7]$ & $[0,U]$ & -- & -- & $0$ & $[0,0.3]$ &  $[0,1]$ & $[0,1]$ & 16028  \\
 $2$ & $50$ & $[0,10]$ & $\frac{3}{2}U-\frac{5}{2}J$ & $[0,0.3U]$ & $[0,3.5]$ & $0$ & $[0,0.35]$ & $[0,1]$ & $[0,1]$ & 18182 \\
  $3$ & $10$ & $[0,7]$ & $[-U,4U]$ & $[0,0.28U]$ & -- & $0.1$ & $[0,0.35]$ & $[0,1]$ & $[0,1]$ & 86431  
\end{tabular}
\caption{Summary of the parameters used to create training datasets for \NNG{}, where the total number of training examples is denoted by $N_{\rm train}$. The $n$-solver utilizes the same dataset as the $3$-orbital \NNG{}. We note that for both the $1$-orbital as well as the $3$-orbital case, extra data is created at half-filling to facilitate the network's task to learn symmetries. In the $3$-orbital case, the DLR coefficients are multiplied by an overall scale of $0.2$. 
QMC simulations were performed using $100000$ CTSEG cycles in the $1$-orbital case, $400000$ CTHYB cycles in the $2$-orbital case and $300000$ CTHYB cycles for $3$ orbitals.}
\label{tab:Gnetparams}
\end{ruledtabular}
\end{table*}

\section{Neural-network details}
\label{app:NN}

 In this section, we explain general details on the creation of the training set and the properties of the NN solvers presented in this work. 
First, we define the mappings represented by the respective NN solvers.
\begin{itemize}
    \item \NNG{} 1-orbital: \\ $
f_{\rm G1}: (\{\Delta(\tau_i)\}, U, \mu_{\rm NN}, \beta) \to \{G(\tau_i)\}$,
    \item \NNG{} 2-orbital: \\ $
f_{\rm G2}: (\{\Delta_{mm}(\tau_i)\}, U, \mu_{\rm NN}, J, \Delta_{\rm cf}) \to \{G_{mm}(\tau_i)\}$,
    \item \NNG{} 3-orbital: \\ $f_{\rm G3}: (\{\Delta_{mm'}(\tau_i)\}, U, \mu_{\rm NN}, J) \to \{G_{mm'}(\tau_i)\}$,
    \item $n$-solver 3-orbital: \\$f_{\rm n}: (\{\Delta_{mm'}(\tau_i)\}, U, \mu_{\rm NN}, J) \to n_{m}$
\end{itemize}

Here, we set 
$\Delta\equiv\Delta_{\uparrow}=\Delta_{\downarrow}$, i.e. assume symmetry with respect to spin flavors. In addition,  we define the effective chemical potential $\mu_{\mathrm{NN}}\equiv -(\epsilon_d-\mu)$, which the neural network takes as input (instead of separate values of $\epsilon_d$ and $\mu$). We note that instead of directly inputting $\mu_{\rm NN}$ into the neural network, we input the ratio $\mu_{\rm NN}/U$.

\subsection{Training dataset}

The general challenge in creating a training dataset consists in creating a
physically meaningful dataset that contains sufficiently distinct hybridization
functions, while at the same time not imposing any information about the
material specifics or the converged solution of a model a-priori. We address
this challenge in the following way. Training examples of the hybridization
function $\Delta$ are created in separate procedures. First, we make use of the
DLR representation (see Section~\ref{app:dlr})
\begin{align}
    \Delta_{m m'}(\tau) \approx \Delta^{{\rm DLR}}_{mm'}(\tau) = \sum_{k=1}^{r} w^k_{m m'} e^{-\omega_k \tau}.
\end{align}
Hybridization functions can now be generated by random choice of the DLR coefficients $w^k_{mm'}$. To ensure that the so-obtained $\Delta$ is physical, it is sufficient to ensure the constraints (i) $w^k\geq 0$ and (ii) $w^k_{mm'}=w^k_{m'm}$. 
 In addition, the training set should balance insulating and metallic solutions. We found that such a balance can be obtained using an exponential envelope to the DLR coefficients, i.e. we choose the coefficients randomly as follows
\begin{align}
w^k_{m m'} &= ((1-\epsilon)\delta_{m m'}+\epsilon)\sqrt{v^k_m v^k_{m'}}, \\ 
v^k_m &= e^{-k\alpha_{m} u^k_{1,m} } u^k_{2,m},
\label{eq:DLRcoeffrand}
\end{align}
where $\delta_{mm'}$ is the Kronecker delta.
Here, the magnitude of $\alpha_{m}$ (and $u^k_{1,m}$) influences if the solution is metallic or insulating. The parameter $\epsilon$ controls the amount of off-diagonal contributions to the hybridization function. 
Here, $\epsilon, \alpha_{m}, u^k_{1,m}, u^k_{2,m}$ are random non-negative numbers chosen in intervals that we specify in Table I, depending on the number of orbitals.  

In the first step of creating the dataset, we generate a set of hybridization functions by the procedure detailed above. We solve the impurity problems defined by these hybridization functions and randomly chosen parameters $U, \mu_{NN}$ (intervals specified below), with the help of CTSEG or CTHYB, and arrive at a set $\mathcal{D_1}$ of input-label pairs. The inputs are $\mathcal{D}_{1, \rm inputs}=(\{ \Delta_{mm'}(\tau_i)\}, U, \mu_{NN}, ...)$ (we do not explicitly write out all input parameters as they depend on the number of orbitals considered)
and the labels are the corresponding Green's function solutions $\mathcal{D}_{1, \rm labels}: = \{G_{mm'}(\tau_i) \}$. The labels are obtained via CTSEG or CTHYB on the input parameters. In the special case of the $n$-solver, the labels are the densities $n_m$.

However, obtaining a well-balanced training data set solely by tuning the
parameters in Eq.~\ref{eq:DLRcoeffrand} is in general a daunting task, and
requires by-hand fine-tuning. We have found that we can balance the training
set in a straightforward way by adding a second step to the training data
generation. In particular, 
we construct a set $\mathcal{D}_2$ that we use to enhance our training dataset. It consists of $n_2$ inputs of the form 
$(\{ \Delta_{mm'}(\tau_i)\}, U, \mu_{NN}, ...)$, where we choose $\{
\Delta_{mm'}(\tau_i)\} \in \mathcal{D}_{1, \rm labels}$. If $U, \mu_{\rm NN},
...$ have the same values as chosen for the set $\mathcal{D}_{1, \rm inputs}$,
this would correspond to the second DMFT-iteration on the Bethe lattice ($\Delta = G$ self-consistency condition). Here,
we include both a set with $U, \mu_{\rm NN}, ...$ to have the same values as in
$\mathcal{D}_{1, \rm inputs}$, as well as new random choices for these
parameters. This procedure has the goal to construct a meaningful "randomness"
for hybridization functions. Then, CTSEG or CTHYB on these inputs creates the
corresponding labels. This procedure can in principle be iterated with new
random values of $U, \mu_{\rm NN} ...$ to create distinct, physically
meaningful hybridization functions, that at the same time do {\it not}
correspond to specific DMFT solutions and are thus agnostic to the model
application. We repeat the procedure $1-2$ times. The total training set is the
combination of all of the so-obtained data.
Table~\ref{tab:Gnetparams} summarizes the intervals chosen for the involved random parameters for each solver, the size of the training set and the computational cost to generate it.

\begin{figure}[b!]
    \centering
    \includegraphics[scale=1.0]{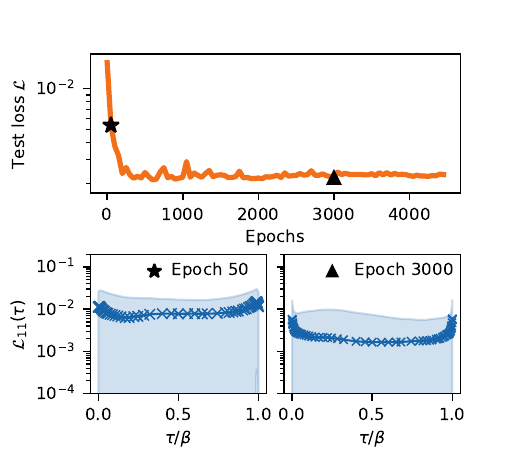}
    \caption{
   Upper panel: Test loss of the $1$-orbital NN solver as a function of number of epochs. Lower panels: Averaged distance as defined in Eg.~(\ref{eq:G_diff_tau}), for epochs $50$ (left) and $3000$ (right). The line and crosses are this averaged distance, the blue shaded regions indicate the standard deviation of this measure over the test set.
}
     \label{fig:test_loss_1orb}
\end{figure}

\begin{figure}[tb]
    \centering
    \includegraphics[scale=1.0]{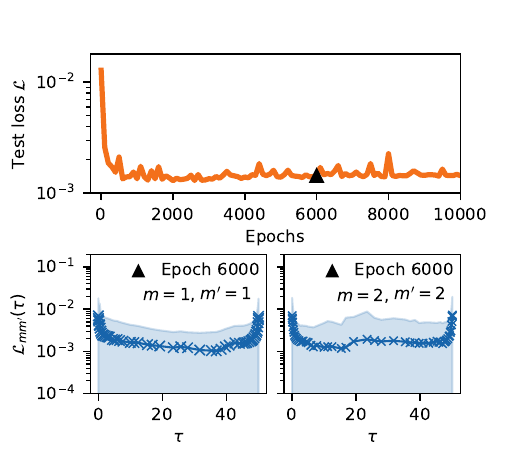}
    \caption{
   Upper panel: Test loss of the $2$-orbital \NNG{} as a function of number of epochs. Lower panels: Averaged distance as defined in Eg.~(\ref{eq:G_diff_tau}), for the epoch $6000$. On the RHS, the averaged distance is plotted for orbital $1$, on the LHS, for orbital $2$. The line and crosses are this averaged distance, the blue shaded regions indicate the standard deviation of this measure over the test set.}
     \label{fig:test_loss_2orb}
\end{figure}

\begin{figure}[tb]
    \centering
    \includegraphics[scale=1.0]{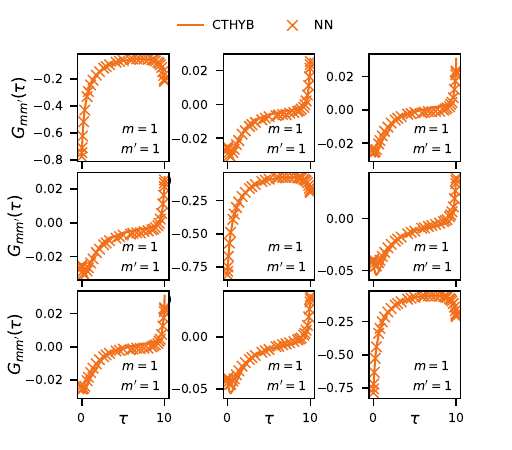}
    \caption{
    Example (single-shot) impurity solution for $U\approx 6.56$, $\mu\approx 0.59$, $J=0.83$, with \NNG{} and CTHYB given the same input $\Delta$ (with non-zero offdiagonal components), for the $3$-orbital case. 
}
     \label{fig:single_shot_3orbs}
\end{figure}

\begin{figure}[tb]
    \centering
    \includegraphics[scale=1.0]{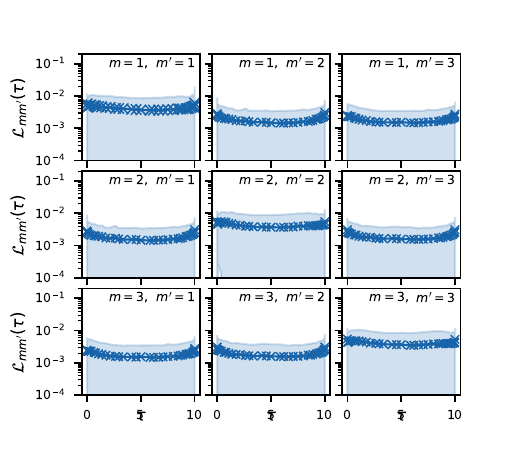}
    \caption{
    Averaged distance as defined in Eg.~(\ref{eq:G_diff_tau}), after training (for the epoch $10000$) in the $3$-orbital case. 
}
     \label{fig:test_loss_3orbs}
\end{figure}

\begin{figure*}[tb]
    \centering
    \includegraphics[scale=1.0]{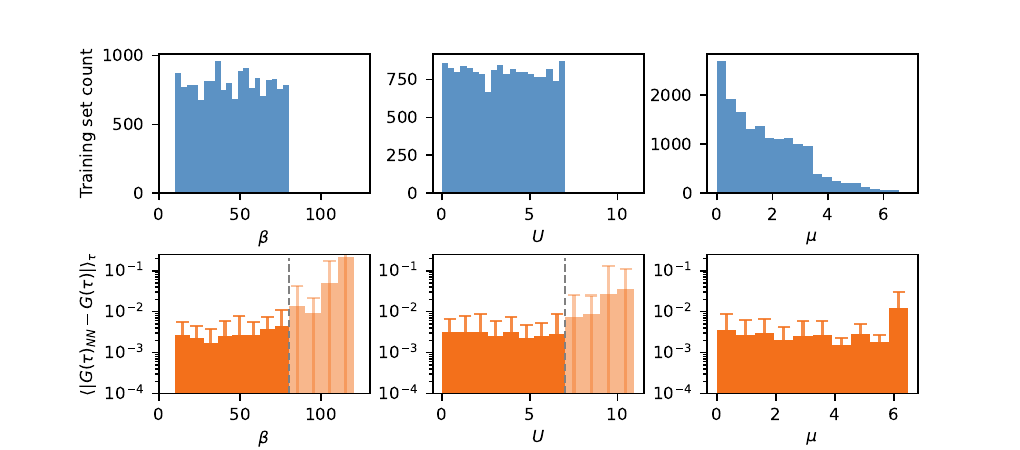}
    \caption{
   Upper panels: Distribution of training examples for the $1$-orbital NN solver over the parameters $\beta$, $U$, $\mu$. Lower panels: Averaged distance $\sum_{\tau} |G_{NN}(\tau)-G(\tau)|$, for epoch $3000$, distributed over parameter values taken from a separate test set. The average is taken over all $\tau$ values. The dashed grey line indicates the parameter boundaries of the training set. We note that the test set utilized for the lower panel is distinct from the ``standard'' test set utilized in Fig.~\ref{fig:test_loss_1orb}, since it includes parameter values of $\beta$ and $U$ outside the trained range.
}
    \label{fig:generalization_1orb}
\end{figure*}

The ranges of $U,\mu_{\rm NN}$ and $J$ are chosen to span physically relevant parameter regimes. The range of $\mu_{\rm NN}$, in particular, is chosen to ensure that the training dataset should cover the whole range of electron occupancies $N=0-2m$ (with exception of the $2$-orbital case, where we trained on half-filling only). The neural network takes the effective chemical potential
$\mu_{\mathrm{NN}}$ as an input, which is randomly sampled as a ratio of $U$, 
\begin{equation}
    \mu_{\mathrm{NN}}/U \in  \left[0,\alpha \right).
\end{equation}
In a simple one-orbital atomic limit, the energies of the $N=1$ and $N=2$ states are given by $E(N=1)=\epsilon_d-\mu$ and $E(N=2)=\epsilon_d-\mu+U$, respectively, therefore $\alpha=1$ can cover all possible occupancies $N=0,1,2$ (see Table~\ref{tab:Gnetparams}). From this relation, half-filling corresponds to $\mu_{\mathrm{eff}}=U/2$. 

For general multi-orbital systems relevant to real materials, the relation between $\mu$ and the electron occupancy $n$ becomes nontrivial, not only because of the multi-orbital Coulomb interaction terms in the Hamiltonian but also the dispersion of the density of states. Nevertheless, the dominant scale controlling charge filling can be set by the ``Hartree" (mean-field) energy with some additional padding for the effect of bandwidth. In the case of $t_{2g}$-orbitals with Kanamori interaction, for example, the Hartree energy, the mean-field energy taking account of only density-density terms, can be written as:
\begin{equation}
    E_{\mathrm{Hartree}} = U\langle n_{m\sigma} \rangle + (U'-J)\sum_{m'}\langle n_{m'\sigma} \rangle + U'\sum_{m'}\langle n_{m'\bar{\sigma}}\rangle .
\end{equation}
In the $M=3$ case, reaching the fully occupied state $N=6$ requires the interaction energy, which yields an upper bound
\begin{equation}
    \mu_{\mathrm{eff,}N=6} = 5(U-2J), 
\end{equation}
which corresponds to the additional interaction energy to add an electron to $N=5$ state. 

In realistic systems, this range of $\mu$ can be further broadened by a finite bandwidth $W$. Taking this into account, and noting that we sample  $J\in\left[ 0, 0.28\right)U$, choosing $\alpha \approx 5$ provides a reasonable coverage of all relevant occupancies from empty to fully-filled states, for the $M=3$ case. In practice, to train the solver for strong correlation phenomena which occurs far away from empty or fully-filled states, we can narrow the sampled $\mu$ range down. Thus, our choice of $\mu_{\mathrm{NN}} \in \left[-U,4U\right)$ safely spans the physically relevant filling range with strong correlation effects. 

\subsection{Architecture} 
{\it \NNG{} 1-orbital.}
We use a Multilayer Perceptron (MLP) with 4 layers and geLU activation function. The input $(\{\Delta(\tau_i\}, U, \mu_{\rm NN}, \beta)$ is flattened into a vector. Crucially, the parameters $U, \mu, \beta$ are in addition embedded using two smaller hidden layers, and the output of these hidden layers is fed into each hidden layer of the main network. This procedure emphasizes the relevance of the parameters $U, \mu, \beta$ and the distinction to the hybridization function. The total number of optimizable parameters is $560289$. 

{\it \NNG{} 2-orbital.}
The architecture follows the same principle as in the $1$-orbital case, but with $5$ instead of $4$ layers, and skip connections. $(\{\Delta_{mm'}(\tau_i)\}, U, \mu_{\rm NN}, J, \Delta_{\rm cf} )$ is flattened into a vector and the parameters $U, \mu_{\rm NN}, J, \Delta_{\rm cf}$ are in addition embedded using $2-3$ smaller hidden layers. The output of this embedding is fed into each hidden layer of the main network. 
The total number of optimizable parameters is $2451834$.

{\it \NNG{} 3-orbital.}
We utilize a residual neural network (ResNet), in order to account for both trainability and increased model size in comparison to the $1$-and $2$-orbital case (the number of optimizable parameters is $17676720$). The input consists of a flattened vector of the quantities $(\{\Delta_{mm'}(\tau_i)\}, U, \mu_{\rm NN}, J) \to \{G_{mm'}(\tau_i)\}$. We note that we did not see improvement in additionally feeding $U, \mu_{\rm NN}, J$ into each layer, as in the case for the $1$-and $2$-orbital solver.

{\it $n$-solver 3-orbital.}
The $n$-solver is constructed by fine-tuning the \NNG{} 3-orbital model. In particular, we add $3$ additional hidden layers to the \NNG{} architecture, with the final output layer of size $3$ (each output neuron corresponding to the density $n_m$ in orbital $m$). These $3$ additional hidden layers contain $125411$ parameters in total. The model is trained by loading the final \NNG{} checkpoint and continuing training with the additional layers. 

\subsection{Training and generalization}
We utilize a mean squared loss for training. In particular, in the case of the \NNG{} solvers, the mean squared loss corresponds to
\begin{align}
\mathcal{L} &= \sum_{m\leq m'}\mathcal{L}_{mm'}, \\ \mathcal{L}_{mm'} &= \frac{1}{|\mathcal{D}|}\sum_{j \in {\rm \mathcal{D}}} \sum_{i} |G^{\rm NN}_{mm'}(\tau_i)-G^{\rm label}_{mm'}(\tau_i)|^2,
\end{align}
where $G^{\rm NN}$ ($G^{\rm label}$) denotes the neural-network output (``ground truth'' Green's function obtained via quantum Monte Carlo). The index $j$ runs over all input - label pairs in the set $\mathcal{D}$. 
We show the loss $\mathcal{L}$ evaluated on a test set $\mathcal{D}_{\rm test}$ as a function of training epochs in the upper panel of Fig.~\ref{fig:test_loss_1orb} for the 1-orbital case, and in the upper panel of Fig.~\ref{fig:test_loss_2orb} for the 2-orbital case.

In addition, we investigate whether the accuracy of the neural-network prediction depends on the value of $\tau$. For this purpose, we plot the imaginary-time dependent mean squared loss, i.e.
\begin{align}
\mathcal{L}_{mm'}(\tau_i) = \frac{1}{|\mathcal{D}|}\sum_{j \in {\rm \mathcal{D}}} |G^{\rm NN}_{mm'}(\tau_i)-G^{\rm label}_{mm'}(\tau_i)|^2.
\label{eq:G_diff_tau}
\end{align}
This quantity is plotted for $\mathcal{D}=\mathcal{D}_{\rm test}$ in the lower panel of Fig~\ref{fig:test_loss_1orb}, early in training (Epoch $50$) and after convergence (Epoch $3000$) in the $1$-orbital case. We find that $G$ is learned to higher precision at intermediate values of $\tau \approx \beta/2$, and has a comparably larger error at the edges $\tau \approx 0, \beta$ (reflecting uncertainties in the density estimation). This potentially stems from typically larger absolute values of variation of the Green's function at $\tau=0$ within physical examples, reflecting different densities. A similar trend is observed in the $2$-orbital case (see lower panel of Fig.~\ref{fig:test_loss_2orb}), where we plot the imaginary-time dependent mean squared loss for the two orbitals separately (i.e. $G_{mm'}$ for $m=m'=1$ and $m=m'=2$) -- most likely enhanced by the fact that the presence of crystal field splitting in the Hamiltonian results in largely differing values at $\tau=0$ and $\tau=\beta$. 

In the $3$-orbital case we allow for non-zero off-diagonal contributions, i.e. $m\neq m'$. We plot both an example Green's function prediction for all components of the Green's function in Fig.~\ref{fig:single_shot_3orbs} as well as the imaginary-time dependent mean squared loss for all components in Fig.~\ref{fig:test_loss_3orbs} after training. Here, the dependence on $\tau$ of the accuracy of the network's prediction is less enhanced than in the $1$-and $2$-orbital case. This may be a result of the larger expressivity of the neural-network model and the larger training set size.

We further probed the generalization abilities of the \NNG{} $1$-orbital solver for different values of $\beta, U, \mu$ in- and outside the range of values the network was trained on. Fig.~\ref{fig:generalization_1orb} shows the result: The upper panel shows the distribution of training examples over values of $\beta, U, \mu$. The lower panel shows the test loss $\mathcal{L}_{mm'}(\tau)$, plotted against values of $\beta, U, \mu$ in an appropriately chosen test set. Values of outside the range $[p_{min}, p_{max}]$ that the network has been trained on (with $p$ corresponding to $\beta$ or $U$) are colored in light color, and $p_{\rm max}$ marked by a grey dashed line. The error bars correspond to the standard deviation of the error in each bin. We find, that the accuracy of the network's prediction is approximately independent of $\beta, U$ and $\mu$, as long as these values are within the respective ranges the model has been trained on. The network is able to generalize outside of this range, but the error increases the further away the parameters are from the original regime.

\begin{figure}[tb]
    \centering
    \includegraphics[width=0.5\textwidth]{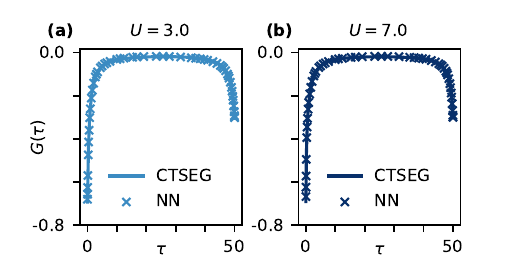}
    \caption{\NNG{} vs CTSEG for impurity occupancy $n=0.6$. (a) $U=3.0$ (b) $U=7.0$}
    \label{fig:n06}
\end{figure}

\begin{figure}[tb]
    \centering
    \includegraphics[width=0.5\textwidth]{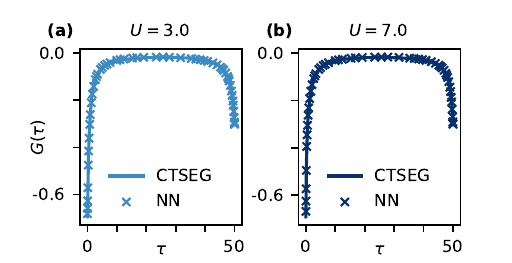}
    \caption{\NNG{} vs CTSEG for impurity occupancy $n=0.8$. (a) $U=3.0$ (b) $U=7.0$}
    \label{fig:n08}
\end{figure}

\section{DLR representation and $\beta$ transferability \label{app:dlr}}

In this work, we use the discrete Lehmann representation (DLR)
expansion~\cite{kaye2022discrete} to compress the imaginary time Green's
function $G(\tau)$ and also take advantage of it to make neural network model
$\beta$-transferable. In this section, we will briefly explain the DLR
expansion for $G(\tau)$ and how we express $G(\tau)$ for different $\beta$
using a single DLR mesh. In particular, we utilize a DLR mesh generated at
fixed $\beta_M$, and then apply it for values of $\beta$ in the range
$(0,\beta_M]$.

Using the Lehmann spectral representation, the imaginary time Green's function can be written as:
\begin{equation}
    G(\tau) = -\int_{-\infty}^{\infty}K_{\beta}(\tau,\omega)\rho(\omega)d\omega \quad \tau \in \left[0,\beta\right]
    \label{eq:g_lehman}
\end{equation}
where $\rho$ is the spectral density and $K_{\beta}$ is the kernel 
\begin{equation}
    K_{\beta}(\tau, \omega) = \frac{e^{-\omega\tau}}{1+e^{-\beta\omega}},
\end{equation}
in the fermionic case. 

If we assume that every spectral density $\rho$ has a finite support 
included in $\left[ -\omega_{\mathrm{max}}, \omega_{\mathrm{max}}\right]$,
we have 
\begin{equation}
   G(\tau) = -\frac{1}{\beta} \int_{-\Lambda}^{\Lambda}K_{\beta=1}(x,\bar \omega)\rho\left(\frac{\bar \omega}{\beta}\right) d\bar \omega 
    \label{eq:DLR_scaled}
\end{equation}
where $x \equiv \frac{\tau}{\beta}$ and  $\Lambda=\beta \omega_{\mathrm{max}}$.
In the DLR representation, the basis only depends on $\Lambda$ and the precision~\cite{kaye2022discrete}.

Equation \eqref{eq:DLR_scaled} shows that we can represent the Green's functions at 
various temperatures as function of $x \in[0,1]$ (hence on the same interval), 
if we take a DLR basis with cut-off $\Lambda = \beta_M  \omega_{\mathrm{max}}$.

The number of DLR basis increases with $\beta_M$. However, in practice, it is
sufficient to consider a logarithmically spaced range of $\beta$, e.g. $\beta
\in [10,100]$. 
In our study, we used the range of $\beta \in \left[10,80 \right]$,
$\omega_{\mathrm{max}}=10$, and $\epsilon=1\mathrm{E}-13$.

To cover larger ranges of temperatures, this procedure could in principle be generated to train separate solvers for various ranges of temperatures - for instance, training of a separate solver for the range $\beta \in \left[80,200 \right]$, with the DLR grid generated at $\beta=200$.

\section{1-orbital model\label{app:1orb}}
\subsection{Non-half-filled cases}
Following the comparison shown in 
Figure 1 of the main text, \NNG{} solver trained for $M=1$ orbital at the range of $\beta\in[10,80]$ is also tested for non-half-filled cases. Figures~\ref{fig:n06} (for $n=0.6$) and ~\ref{fig:n08} (for $n=0.8$) show the comparison between \NNG{} and CTSEG solvers for small ($U=3$) and large ($U=7$) $U$ vales, where $n=1$ corresponds to half-filled case. The calculations was done at $\beta=50$, and again, the \NNG{} solver shows an excellent agreement to CTSEG solver results.

\begin{figure}[tb]
    \centering
    \includegraphics[width=0.5\textwidth]{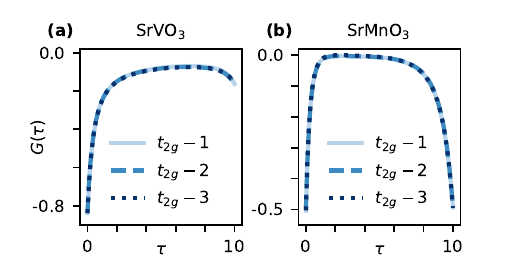}
    \caption{NN Green's function shown for all 3 $t_{2g}$ orbitals in the case of SrVO$_3$ (a) and SrMnO$_3$ (b).}
    \label{fig:check-3orb}
\end{figure}

\section{3-orbital Materials \label{app:3orb}}

\begin{figure}[b!]
    \centering
    \includegraphics[width=0.5\textwidth]{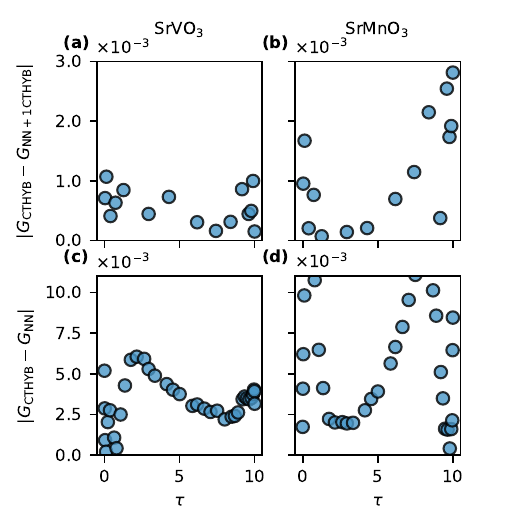}
    \caption{Difference in $G(\tau)$. $|G_{\mathrm{CTHYB}}-G_{\mathrm{NN+1 CTHYB}}|$ for SrVO$_3$ (a) and SrMnO$_3$ (b). $|G_{\mathrm{CTHYB}}-G_{\mathrm{NN}}|$ for SrVO$_3$ (c) and SrMnO$_3$ (d).}
    \label{fig:delG}
\end{figure}

\begin{figure}[tb]
    \centering
    \includegraphics[width=0.5\textwidth]{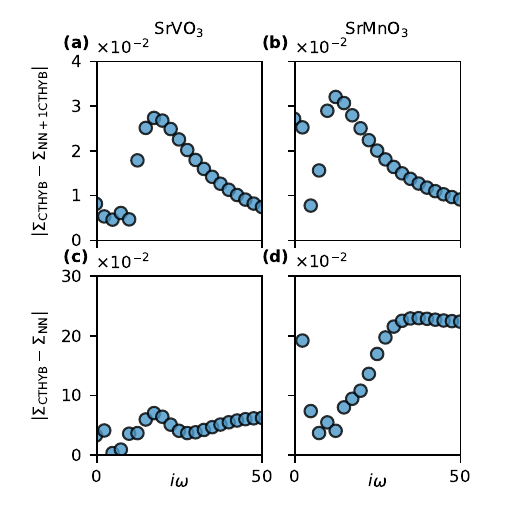}
    \caption{Difference in $\Sigma(i\omega)$. $|\Sigma_{\mathrm{CTHYB}}-\Sigma_{\mathrm{NN+1 CTHYB}}|$ for SrVO$_3$ (a) and SrMnO$_3$ (b). $|\Sigma_{\mathrm{CTHYB}}-\Sigma_{\mathrm{NN}}|$ for SrVO$_3$ (c) and SrMnO$_3$ (d).}
    \label{fig:delS}
\end{figure}

The \NNG{} solver does not impose any symmetry constraints in general multi-orbital problems. In other words, both the input hybridization function and the output Green's function are $M\times M$ (ignoring spin) block matrices, whose diagonal blocks can differ from each other and whose off-diagonal blocks can be non-vanishing. The non-degenerate two-orbital system with crystal-field splitting presented in the main text is an example of a case with different diagonal blocks.
This absence of enforced symmetry is in fact a general property shared by conventional impurity solvers such as CTQMC.

It is therefore essential that the \NNG{} solver be able to predict the Green's function with the symmetry expected from the symmetry of the local Hamiltonian and the interaction Hamiltonian. In other words, while the solver does not assume any symmetry a priori, it has to predict the Green's function with the physically required symmetry whenever it is present in the problem, up to numerical error.

For the real material examples considered in this work, SrVO$_3$ and SrMnO$_3$, the local octahedral ligand field enforces a threefold degeneracy of $t_{2g}$ orbitals of an impurity problem. Consequently, the (DFT) hybridization function is block diagonal (in principle, and in practice as well) with vanishing off-block-diagonal components, and the Green's function solution obtained from self-consistent DMFT loop is also expected to preserve this symmetry. In Fig.~\ref{fig:check-3orb}, we show that the \NNG{} solver correctly reproduces this expected symmetric structure of the Green's function for both SrVO$_3$ and SrMnO$_3$. 

Further, we show the accuracy of the \NNG{} prediction of the Green's function and the self-energy in Figs.~\ref{fig:delG} and \ref{fig:delS}. In particular, we plot the difference to reference CTHYB calculations, and compare both the NN-accelerated strategy as well as solely using \NNG{}. 


\begin{thebibliography}{32}%
\makeatletter
\providecommand \@ifxundefined [1]{%
 \@ifx{#1\undefined}
}%
\providecommand \@ifnum [1]{%
 \ifnum #1\expandafter \@firstoftwo
 \else \expandafter \@secondoftwo
 \fi
}%
\providecommand \@ifx [1]{%
 \ifx #1\expandafter \@firstoftwo
 \else \expandafter \@secondoftwo
 \fi
}%
\providecommand \natexlab [1]{#1}%
\providecommand \enquote  [1]{``#1''}%
\providecommand \bibnamefont  [1]{#1}%
\providecommand \bibfnamefont [1]{#1}%
\providecommand \citenamefont [1]{#1}%
\providecommand \href@noop [0]{\@secondoftwo}%
\providecommand \href [0]{\begingroup \@sanitize@url \@href}%
\providecommand \@href[1]{\@@startlink{#1}\@@href}%
\providecommand \@@href[1]{\endgroup#1\@@endlink}%
\providecommand \@sanitize@url [0]{\catcode `\\12\catcode `\$12\catcode
  `\&12\catcode `\#12\catcode `\^12\catcode `\_12\catcode `\%12\relax}%
\providecommand \@@startlink[1]{}%
\providecommand \@@endlink[0]{}%
\providecommand \url  [0]{\begingroup\@sanitize@url \@url }%
\providecommand \@url [1]{\endgroup\@href {#1}{\urlprefix }}%
\providecommand \urlprefix  [0]{URL }%
\providecommand \Eprint [0]{\href }%
\providecommand \doibase [0]{https://doi.org/}%
\providecommand \selectlanguage [0]{\@gobble}%
\providecommand \bibinfo  [0]{\@secondoftwo}%
\providecommand \bibfield  [0]{\@secondoftwo}%
\providecommand \translation [1]{[#1]}%
\providecommand \BibitemOpen [0]{}%
\providecommand \bibitemStop [0]{}%
\providecommand \bibitemNoStop [0]{.\EOS\space}%
\providecommand \EOS [0]{\spacefactor3000\relax}%
\providecommand \BibitemShut  [1]{\csname bibitem#1\endcsname}%
\let\auto@bib@innerbib\@empty
\bibitem [{\citenamefont {Dirac}(1929)}]{dirac1929}%
  \BibitemOpen
  \bibfield  {author} {\bibinfo {author} {\bibfnamefont {P.~A.~M.}\
  \bibnamefont {Dirac}},\ }\href {https://doi.org/10.1098/rspa.1929.0094}
  {\bibfield  {journal} {\bibinfo  {journal} {Proceedings of the Royal Society
  of London. Series A, Containing Papers of a Mathematical and Physical
  Character}\ }\textbf {\bibinfo {volume} {123}},\ \bibinfo {pages} {714}
  (\bibinfo {year} {1929})}\BibitemShut {NoStop}%
\bibitem [{\citenamefont {Georges}\ and\ \citenamefont
  {Kotliar}(1992)}]{georges1992hubbard}%
  \BibitemOpen
  \bibfield  {author} {\bibinfo {author} {\bibfnamefont {A.}~\bibnamefont
  {Georges}}\ and\ \bibinfo {author} {\bibfnamefont {G.}~\bibnamefont
  {Kotliar}},\ }\href
  {https://doi.org/https://doi.org/10.1103/PhysRevB.45.6479} {\bibfield
  {journal} {\bibinfo  {journal} {Phys. Rev. B}\ }\textbf {\bibinfo {volume}
  {45}},\ \bibinfo {pages} {6479} (\bibinfo {year} {1992})}\BibitemShut
  {NoStop}%
\bibitem [{\citenamefont {Georges}\ \emph {et~al.}(1996)\citenamefont
  {Georges}, \citenamefont {Kotliar}, \citenamefont {Krauth},\ and\
  \citenamefont {Rozenberg}}]{georgesRMP1996}%
  \BibitemOpen
  \bibfield  {author} {\bibinfo {author} {\bibfnamefont {A.}~\bibnamefont
  {Georges}}, \bibinfo {author} {\bibfnamefont {G.}~\bibnamefont {Kotliar}},
  \bibinfo {author} {\bibfnamefont {W.}~\bibnamefont {Krauth}},\ and\ \bibinfo
  {author} {\bibfnamefont {M.~J.}\ \bibnamefont {Rozenberg}},\ }\href
  {https://doi.org/10.1103/RevModPhys.68.13} {\bibfield  {journal} {\bibinfo
  {journal} {Rev. Mod. Phys.}\ }\textbf {\bibinfo {volume} {68}},\ \bibinfo
  {pages} {13} (\bibinfo {year} {1996})}\BibitemShut {NoStop}%
\bibitem [{\citenamefont {Kotliar}\ \emph {et~al.}(2006)\citenamefont
  {Kotliar}, \citenamefont {Savrasov}, \citenamefont {Haule}, \citenamefont
  {Oudovenko}, \citenamefont {Parcollet},\ and\ \citenamefont
  {Marianetti}}]{kotliarRMP2006}%
  \BibitemOpen
  \bibfield  {author} {\bibinfo {author} {\bibfnamefont {G.}~\bibnamefont
  {Kotliar}}, \bibinfo {author} {\bibfnamefont {S.~Y.}\ \bibnamefont
  {Savrasov}}, \bibinfo {author} {\bibfnamefont {K.}~\bibnamefont {Haule}},
  \bibinfo {author} {\bibfnamefont {V.~S.}\ \bibnamefont {Oudovenko}}, \bibinfo
  {author} {\bibfnamefont {O.}~\bibnamefont {Parcollet}},\ and\ \bibinfo
  {author} {\bibfnamefont {C.}~\bibnamefont {Marianetti}},\ }\href
  {https://doi.org/10.1103/RevModPhys.78.865} {\bibfield  {journal} {\bibinfo
  {journal} {Rev. Mod. Phys.}\ }\textbf {\bibinfo {volume} {78}},\ \bibinfo
  {pages} {865} (\bibinfo {year} {2006})}\BibitemShut {NoStop}%
\bibitem [{\citenamefont {Gull}\ \emph {et~al.}(2011)\citenamefont {Gull},
  \citenamefont {Millis}, \citenamefont {Lichtenstein}, \citenamefont
  {Rubtsov}, \citenamefont {Troyer},\ and\ \citenamefont
  {Werner}}]{gull2011continuous}%
  \BibitemOpen
  \bibfield  {author} {\bibinfo {author} {\bibfnamefont {E.}~\bibnamefont
  {Gull}}, \bibinfo {author} {\bibfnamefont {A.~J.}\ \bibnamefont {Millis}},
  \bibinfo {author} {\bibfnamefont {A.~I.}\ \bibnamefont {Lichtenstein}},
  \bibinfo {author} {\bibfnamefont {A.~N.}\ \bibnamefont {Rubtsov}}, \bibinfo
  {author} {\bibfnamefont {M.}~\bibnamefont {Troyer}},\ and\ \bibinfo {author}
  {\bibfnamefont {P.}~\bibnamefont {Werner}},\ }\href
  {https://doi.org/https://doi.org/10.1103/RevModPhys.83.349} {\bibfield
  {journal} {\bibinfo  {journal} {Rev. Mod. Phys.}\ }\textbf {\bibinfo {volume}
  {83}},\ \bibinfo {pages} {349} (\bibinfo {year} {2011})}\BibitemShut
  {NoStop}%
\bibitem [{\citenamefont {Werner}\ \emph {et~al.}(2006)\citenamefont {Werner},
  \citenamefont {Comanac}, \citenamefont {De’Medici}, \citenamefont
  {Troyer},\ and\ \citenamefont {Millis}}]{werner2006continuous}%
  \BibitemOpen
  \bibfield  {author} {\bibinfo {author} {\bibfnamefont {P.}~\bibnamefont
  {Werner}}, \bibinfo {author} {\bibfnamefont {A.}~\bibnamefont {Comanac}},
  \bibinfo {author} {\bibfnamefont {L.}~\bibnamefont {De’Medici}}, \bibinfo
  {author} {\bibfnamefont {M.}~\bibnamefont {Troyer}},\ and\ \bibinfo {author}
  {\bibfnamefont {A.~J.}\ \bibnamefont {Millis}},\ }\href
  {https://doi.org/https://doi.org/10.1103/PhysRevLett.97.076405} {\bibfield
  {journal} {\bibinfo  {journal} {Phys. Rev. Lett.}\ }\textbf {\bibinfo
  {volume} {97}},\ \bibinfo {pages} {076405} (\bibinfo {year}
  {2006})}\BibitemShut {NoStop}%
\bibitem [{\citenamefont {Werner}\ and\ \citenamefont
  {Millis}(2006)}]{werner2006hybridization}%
  \BibitemOpen
  \bibfield  {author} {\bibinfo {author} {\bibfnamefont {P.}~\bibnamefont
  {Werner}}\ and\ \bibinfo {author} {\bibfnamefont {A.~J.}\ \bibnamefont
  {Millis}},\ }\href {https://doi.org/doi.org/10.1103/PhysRevB.74.155107}
  {\bibfield  {journal} {\bibinfo  {journal} {Phys. Rev. B—Condensed Matter
  and Materials Physics}\ }\textbf {\bibinfo {volume} {74}},\ \bibinfo {pages}
  {155107} (\bibinfo {year} {2006})}\BibitemShut {NoStop}%
\bibitem [{\citenamefont {Rubtsov}\ \emph {et~al.}(2005)\citenamefont
  {Rubtsov}, \citenamefont {Savkin},\ and\ \citenamefont
  {Lichtenstein}}]{rubtsov2005continuous}%
  \BibitemOpen
  \bibfield  {author} {\bibinfo {author} {\bibfnamefont {A.~N.}\ \bibnamefont
  {Rubtsov}}, \bibinfo {author} {\bibfnamefont {V.~V.}\ \bibnamefont
  {Savkin}},\ and\ \bibinfo {author} {\bibfnamefont {A.~I.}\ \bibnamefont
  {Lichtenstein}},\ }\href
  {https://doi.org/https://doi.org/10.1103/PhysRevB.72.035122} {\bibfield
  {journal} {\bibinfo  {journal} {Phys. Rev. B—Condensed Matter and Materials
  Physics}\ }\textbf {\bibinfo {volume} {72}},\ \bibinfo {pages} {035122}
  (\bibinfo {year} {2005})}\BibitemShut {NoStop}%
\bibitem [{\citenamefont {Gull}\ \emph {et~al.}(2008)\citenamefont {Gull},
  \citenamefont {Werner}, \citenamefont {Parcollet},\ and\ \citenamefont
  {Troyer}}]{gull2008continuous}%
  \BibitemOpen
  \bibfield  {author} {\bibinfo {author} {\bibfnamefont {E.}~\bibnamefont
  {Gull}}, \bibinfo {author} {\bibfnamefont {P.}~\bibnamefont {Werner}},
  \bibinfo {author} {\bibfnamefont {O.}~\bibnamefont {Parcollet}},\ and\
  \bibinfo {author} {\bibfnamefont {M.}~\bibnamefont {Troyer}},\ }\href
  {https://doi.org/10.1209/0295-5075/82/57003} {\bibfield  {journal} {\bibinfo
  {journal} {Europhys. Lett.}\ }\textbf {\bibinfo {volume} {82}},\ \bibinfo
  {pages} {57003} (\bibinfo {year} {2008})}\BibitemShut {NoStop}%
\bibitem [{\citenamefont {Bulla}\ \emph {et~al.}(2008)\citenamefont {Bulla},
  \citenamefont {Costi},\ and\ \citenamefont {Pruschke}}]{bulla2008numerical}%
  \BibitemOpen
  \bibfield  {author} {\bibinfo {author} {\bibfnamefont {R.}~\bibnamefont
  {Bulla}}, \bibinfo {author} {\bibfnamefont {T.~A.}\ \bibnamefont {Costi}},\
  and\ \bibinfo {author} {\bibfnamefont {T.}~\bibnamefont {Pruschke}},\ }\href
  {https://doi.org/https://doi.org/10.1103/RevModPhys.80.395} {\bibfield
  {journal} {\bibinfo  {journal} {Rev. Mod. Phys.}\ }\textbf {\bibinfo {volume}
  {80}},\ \bibinfo {pages} {395} (\bibinfo {year} {2008})}\BibitemShut
  {NoStop}%
\bibitem [{\citenamefont {Bulla}(1999)}]{bulla1999zero}%
  \BibitemOpen
  \bibfield  {author} {\bibinfo {author} {\bibfnamefont {R.}~\bibnamefont
  {Bulla}},\ }\href
  {https://doi.org/https://doi.org/10.1103/PhysRevLett.83.136} {\bibfield
  {journal} {\bibinfo  {journal} {Phys. Rev. Lett.}\ }\textbf {\bibinfo
  {volume} {83}},\ \bibinfo {pages} {136} (\bibinfo {year} {1999})}\BibitemShut
  {NoStop}%
\bibitem [{\citenamefont {White}(1992)}]{white1992density}%
  \BibitemOpen
  \bibfield  {author} {\bibinfo {author} {\bibfnamefont {S.~R.}\ \bibnamefont
  {White}},\ }\href
  {https://doi.org/https://doi.org/10.1103/PhysRevLett.69.2863} {\bibfield
  {journal} {\bibinfo  {journal} {Phys. Rev. Lett.}\ }\textbf {\bibinfo
  {volume} {69}},\ \bibinfo {pages} {2863} (\bibinfo {year}
  {1992})}\BibitemShut {NoStop}%
\bibitem [{\citenamefont {Schollw{\"o}ck}(2005)}]{schollwock2005density}%
  \BibitemOpen
  \bibfield  {author} {\bibinfo {author} {\bibfnamefont {U.}~\bibnamefont
  {Schollw{\"o}ck}},\ }\href
  {https://doi.org/https://doi.org/10.1103/RevModPhys.77.259} {\bibfield
  {journal} {\bibinfo  {journal} {Rev. Mod. Phys.}\ }\textbf {\bibinfo {volume}
  {77}},\ \bibinfo {pages} {259} (\bibinfo {year} {2005})}\BibitemShut
  {NoStop}%
\bibitem [{\citenamefont {{\"O}stlund}\ and\ \citenamefont
  {Rommer}(1995)}]{ostlund1995thermodynamic}%
  \BibitemOpen
  \bibfield  {author} {\bibinfo {author} {\bibfnamefont {S.}~\bibnamefont
  {{\"O}stlund}}\ and\ \bibinfo {author} {\bibfnamefont {S.}~\bibnamefont
  {Rommer}},\ }\href
  {https://doi.org/https://doi.org/10.1103/PhysRevLett.75.3537} {\bibfield
  {journal} {\bibinfo  {journal} {Phys. Rev. Lett.}\ }\textbf {\bibinfo
  {volume} {75}},\ \bibinfo {pages} {3537} (\bibinfo {year}
  {1995})}\BibitemShut {NoStop}%
\bibitem [{\citenamefont {Anisimov}\ \emph {et~al.}(1997)\citenamefont
  {Anisimov}, \citenamefont {Poteryaev}, \citenamefont {Korotin}, \citenamefont
  {Anokhin},\ and\ \citenamefont {Kotliar}}]{anisimov1997first}%
  \BibitemOpen
  \bibfield  {author} {\bibinfo {author} {\bibfnamefont {V.}~\bibnamefont
  {Anisimov}}, \bibinfo {author} {\bibfnamefont {A.}~\bibnamefont {Poteryaev}},
  \bibinfo {author} {\bibfnamefont {M.}~\bibnamefont {Korotin}}, \bibinfo
  {author} {\bibfnamefont {A.}~\bibnamefont {Anokhin}},\ and\ \bibinfo {author}
  {\bibfnamefont {G.}~\bibnamefont {Kotliar}},\ }\href
  {https://doi.org/10.1088/0953-8984/9/35/010} {\bibfield  {journal} {\bibinfo
  {journal} {J. Phys. Condens. Matter.}\ }\textbf {\bibinfo {volume} {9}},\
  \bibinfo {pages} {7359} (\bibinfo {year} {1997})}\BibitemShut {NoStop}%
\bibitem [{\citenamefont {Lichtenstein}\ and\ \citenamefont
  {Katsnelson}(1998)}]{lichtenstein1998}%
  \BibitemOpen
  \bibfield  {author} {\bibinfo {author} {\bibfnamefont {A.~I.}\ \bibnamefont
  {Lichtenstein}}\ and\ \bibinfo {author} {\bibfnamefont {M.~I.}\ \bibnamefont
  {Katsnelson}},\ }\href {https://doi.org/10.1103/PhysRevB.57.6884} {\bibfield
  {journal} {\bibinfo  {journal} {Phys. Rev. B}\ }\textbf {\bibinfo {volume}
  {57}},\ \bibinfo {pages} {6884} (\bibinfo {year} {1998})}\BibitemShut
  {NoStop}%
\bibitem [{\citenamefont {Arsenault}\ \emph {et~al.}(2014)\citenamefont
  {Arsenault}, \citenamefont {{Lopez-Bezanilla}}, \citenamefont
  {Von~Lilienfeld},\ and\ \citenamefont {Millis}}]{arsenault2014machine}%
  \BibitemOpen
  \bibfield  {author} {\bibinfo {author} {\bibfnamefont {L.-F.}\ \bibnamefont
  {Arsenault}}, \bibinfo {author} {\bibfnamefont {A.}~\bibnamefont
  {{Lopez-Bezanilla}}}, \bibinfo {author} {\bibfnamefont {O.~A.}\ \bibnamefont
  {Von~Lilienfeld}},\ and\ \bibinfo {author} {\bibfnamefont {A.~J.}\
  \bibnamefont {Millis}},\ }\href {https://doi.org/10.1103/PhysRevB.90.155136}
  {\bibfield  {journal} {\bibinfo  {journal} {Phys. Rev. B}\ }\textbf {\bibinfo
  {volume} {90}},\ \bibinfo {pages} {155136} (\bibinfo {year}
  {2014})}\BibitemShut {NoStop}%
\bibitem [{\citenamefont {Sheridan}\ \emph {et~al.}(2021)\citenamefont
  {Sheridan}, \citenamefont {Rhodes}, \citenamefont {Jamet}, \citenamefont
  {Rungger},\ and\ \citenamefont {Weber}}]{sheridan2021data}%
  \BibitemOpen
  \bibfield  {author} {\bibinfo {author} {\bibfnamefont {E.}~\bibnamefont
  {Sheridan}}, \bibinfo {author} {\bibfnamefont {C.}~\bibnamefont {Rhodes}},
  \bibinfo {author} {\bibfnamefont {F.}~\bibnamefont {Jamet}}, \bibinfo
  {author} {\bibfnamefont {I.}~\bibnamefont {Rungger}},\ and\ \bibinfo {author}
  {\bibfnamefont {C.}~\bibnamefont {Weber}},\ }\href
  {https://doi.org/10.1103/PhysRevB.104.205120} {\bibfield  {journal} {\bibinfo
   {journal} {Phys. Rev. B}\ }\textbf {\bibinfo {volume} {104}},\ \bibinfo
  {pages} {205120} (\bibinfo {year} {2021})}\BibitemShut {NoStop}%
\bibitem [{\citenamefont {Sturm}\ \emph {et~al.}(2021)\citenamefont {Sturm},
  \citenamefont {Carbone}, \citenamefont {Lu}, \citenamefont {Weichselbaum},\
  and\ \citenamefont {Konik}}]{sturm2021predicting}%
  \BibitemOpen
  \bibfield  {author} {\bibinfo {author} {\bibfnamefont {E.~J.}\ \bibnamefont
  {Sturm}}, \bibinfo {author} {\bibfnamefont {M.~R.}\ \bibnamefont {Carbone}},
  \bibinfo {author} {\bibfnamefont {D.}~\bibnamefont {Lu}}, \bibinfo {author}
  {\bibfnamefont {A.}~\bibnamefont {Weichselbaum}},\ and\ \bibinfo {author}
  {\bibfnamefont {R.~M.}\ \bibnamefont {Konik}},\ }\href
  {https://doi.org/10.1103/PhysRevB.103.245118} {\bibfield  {journal} {\bibinfo
   {journal} {Phys. Rev. B}\ }\textbf {\bibinfo {volume} {103}},\ \bibinfo
  {pages} {245118} (\bibinfo {year} {2021})}\BibitemShut {NoStop}%
\bibitem [{\citenamefont {Agapov}\ \emph {et~al.}(2024)\citenamefont {Agapov},
  \citenamefont {Bertomeu}, \citenamefont {Carballo}, \citenamefont {Mendl},\
  and\ \citenamefont {Sander}}]{agapov2024predicting}%
  \BibitemOpen
  \bibfield  {author} {\bibinfo {author} {\bibfnamefont {E.}~\bibnamefont
  {Agapov}}, \bibinfo {author} {\bibfnamefont {O.}~\bibnamefont {Bertomeu}},
  \bibinfo {author} {\bibfnamefont {A.}~\bibnamefont {Carballo}}, \bibinfo
  {author} {\bibfnamefont {C.~B.}\ \bibnamefont {Mendl}},\ and\ \bibinfo
  {author} {\bibfnamefont {A.}~\bibnamefont {Sander}},\ }\bibfield  {journal}
  {\bibinfo  {journal} {arXiv:2411.13644}\ }\href
  {https://arxiv.org/abs/2411.13644} {} (\bibinfo {year} {2024})\BibitemShut
  {NoStop}%
\bibitem [{\citenamefont {Lee}\ \emph {et~al.}(2025)\citenamefont {Lee},
  \citenamefont {Zhao}, \citenamefont {Booth}, \citenamefont {Ge},\ and\
  \citenamefont {Weber}}]{lee2025language}%
  \BibitemOpen
  \bibfield  {author} {\bibinfo {author} {\bibfnamefont {H.}~\bibnamefont
  {Lee}}, \bibinfo {author} {\bibfnamefont {Z.}~\bibnamefont {Zhao}}, \bibinfo
  {author} {\bibfnamefont {G.~H.}\ \bibnamefont {Booth}}, \bibinfo {author}
  {\bibfnamefont {W.}~\bibnamefont {Ge}},\ and\ \bibinfo {author}
  {\bibfnamefont {C.}~\bibnamefont {Weber}},\ }\href
  {https://doi.org/10.1103/bk5q-pfb2} {\bibfield  {journal} {\bibinfo
  {journal} {Phys. Rev. B}\ }\textbf {\bibinfo {volume} {112}},\ \bibinfo
  {pages} {035165} (\bibinfo {year} {2025})}\BibitemShut {NoStop}%
\bibitem [{\citenamefont {Mitra}\ and\ \citenamefont
  {Banerjee}(2025)}]{mitra2025deep}%
  \BibitemOpen
  \bibfield  {author} {\bibinfo {author} {\bibfnamefont {P.}~\bibnamefont
  {Mitra}}\ and\ \bibinfo {author} {\bibfnamefont {H.}~\bibnamefont
  {Banerjee}},\ }\href
  {https://chemrxiv.org/doi/full/10.26434/chemrxiv-2025-dp7rd} {\bibfield
  {journal} {\bibinfo  {journal} {chemrxiv-2025-dp7rd}\ } (\bibinfo {year}
  {2025})}\BibitemShut {NoStop}%
\bibitem [{\citenamefont {Rao}\ and\ \citenamefont {Zhu}(2026)}]{rao2026}%
  \BibitemOpen
  \bibfield  {author} {\bibinfo {author} {\bibfnamefont {R.}~\bibnamefont
  {Rao}}\ and\ \bibinfo {author} {\bibfnamefont {L.}~\bibnamefont {Zhu}},\
  }\bibfield  {journal} {\bibinfo  {journal} {arXiv:2512.25061}\ }\href@noop {}
  {} (\bibinfo {year} {2026})\BibitemShut {NoStop}%
\bibitem [{\citenamefont {Kaye}\ \emph {et~al.}(2022)\citenamefont {Kaye},
  \citenamefont {Chen},\ and\ \citenamefont {Parcollet}}]{kaye2022discrete}%
  \BibitemOpen
  \bibfield  {author} {\bibinfo {author} {\bibfnamefont {J.}~\bibnamefont
  {Kaye}}, \bibinfo {author} {\bibfnamefont {K.}~\bibnamefont {Chen}},\ and\
  \bibinfo {author} {\bibfnamefont {O.}~\bibnamefont {Parcollet}},\ }\href
  {https://doi.org/https://doi.org/10.1103/PhysRevB.105.235115} {\bibfield
  {journal} {\bibinfo  {journal} {Phys. Rev. B}\ }\textbf {\bibinfo {volume}
  {105}},\ \bibinfo {pages} {235115} (\bibinfo {year} {2022})}\BibitemShut
  {NoStop}%
\bibitem [{Note1()}]{Note1}%
  \BibitemOpen
  \bibinfo {note} {Alternatively, this can be seen as the exact solution of a
  fully-connected Hubbard model with random hopping, in which case no magnetic
  phase is present\cite {georgesRMP1996}.}\BibitemShut {Stop}%
\bibitem [{\citenamefont {Kanamori}(1963)}]{kanamori1963electron}%
  \BibitemOpen
  \bibfield  {author} {\bibinfo {author} {\bibfnamefont {J.}~\bibnamefont
  {Kanamori}},\ }\href {https://doi.org/doi.org/10.1143/PTP.30.275} {\bibfield
  {journal} {\bibinfo  {journal} {Prog. Theor. Phys.}\ }\textbf {\bibinfo
  {volume} {30}},\ \bibinfo {pages} {275} (\bibinfo {year} {1963})}\BibitemShut
  {NoStop}%
\bibitem [{\citenamefont {Georges}\ \emph {et~al.}(2013)\citenamefont
  {Georges}, \citenamefont {Medici},\ and\ \citenamefont
  {Mravlje}}]{georges2013strong}%
  \BibitemOpen
  \bibfield  {author} {\bibinfo {author} {\bibfnamefont {A.}~\bibnamefont
  {Georges}}, \bibinfo {author} {\bibfnamefont {L.~d.}\ \bibnamefont
  {Medici}},\ and\ \bibinfo {author} {\bibfnamefont {J.}~\bibnamefont
  {Mravlje}},\ }\href
  {https://doi.org/doi.org/10.1146/annurev-conmatphys-020911-125045} {\bibfield
   {journal} {\bibinfo  {journal} {Annu. Rev. Condens. Matter Phys.}\ }\textbf
  {\bibinfo {volume} {4}},\ \bibinfo {pages} {137} (\bibinfo {year}
  {2013})}\BibitemShut {NoStop}%
\bibitem [{\citenamefont {Werner}\ and\ \citenamefont
  {Millis}(2007)}]{werner2007high}%
  \BibitemOpen
  \bibfield  {author} {\bibinfo {author} {\bibfnamefont {P.}~\bibnamefont
  {Werner}}\ and\ \bibinfo {author} {\bibfnamefont {A.~J.}\ \bibnamefont
  {Millis}},\ }\href
  {https://doi.org/https://doi.org/10.1103/PhysRevLett.99.126405} {\bibfield
  {journal} {\bibinfo  {journal} {Phys. Rev. Lett.}\ }\textbf {\bibinfo
  {volume} {99}},\ \bibinfo {pages} {126405} (\bibinfo {year}
  {2007})}\BibitemShut {NoStop}%
\bibitem [{\citenamefont {Fujimori}\ \emph {et~al.}(1992)\citenamefont
  {Fujimori}, \citenamefont {Hase}, \citenamefont {Namatame}, \citenamefont
  {Fujishima}, \citenamefont {Tokura}, \citenamefont {Eisaki}, \citenamefont
  {Uchida}, \citenamefont {Takegahara},\ and\ \citenamefont
  {De~Groot}}]{fujimori1992evolution}%
  \BibitemOpen
  \bibfield  {author} {\bibinfo {author} {\bibfnamefont {A.}~\bibnamefont
  {Fujimori}}, \bibinfo {author} {\bibfnamefont {I.}~\bibnamefont {Hase}},
  \bibinfo {author} {\bibfnamefont {H.}~\bibnamefont {Namatame}}, \bibinfo
  {author} {\bibfnamefont {Y.}~\bibnamefont {Fujishima}}, \bibinfo {author}
  {\bibfnamefont {Y.}~\bibnamefont {Tokura}}, \bibinfo {author} {\bibfnamefont
  {H.}~\bibnamefont {Eisaki}}, \bibinfo {author} {\bibfnamefont
  {S.}~\bibnamefont {Uchida}}, \bibinfo {author} {\bibfnamefont
  {K.}~\bibnamefont {Takegahara}},\ and\ \bibinfo {author} {\bibfnamefont
  {F.}~\bibnamefont {De~Groot}},\ }\href
  {https://doi.org/https://doi.org/10.1103/PhysRevLett.69.1796} {\bibfield
  {journal} {\bibinfo  {journal} {Phys. Rev. Lett.}\ }\textbf {\bibinfo
  {volume} {69}},\ \bibinfo {pages} {1796} (\bibinfo {year}
  {1992})}\BibitemShut {NoStop}%
\bibitem [{\citenamefont {Lee}\ and\ \citenamefont
  {Iguchi}(1995)}]{lee1995electronic}%
  \BibitemOpen
  \bibfield  {author} {\bibinfo {author} {\bibfnamefont {K.}~\bibnamefont
  {Lee}}\ and\ \bibinfo {author} {\bibfnamefont {E.}~\bibnamefont {Iguchi}},\
  }\href {https://doi.org/https://doi.org/10.1006/jssc.1995.1035} {\bibfield
  {journal} {\bibinfo  {journal} {J. Solid State Chem.}\ }\textbf {\bibinfo
  {volume} {114}},\ \bibinfo {pages} {242} (\bibinfo {year}
  {1995})}\BibitemShut {NoStop}%
\bibitem [{\citenamefont {Kim}\ \emph {et~al.}(2010)\citenamefont {Kim},
  \citenamefont {Lee}, \citenamefont {Dabrowski}, \citenamefont {Kolesnik},
  \citenamefont {Lee}, \citenamefont {Kim}, \citenamefont {Min},\ and\
  \citenamefont {Kang}}]{kim2010photoemission}%
  \BibitemOpen
  \bibfield  {author} {\bibinfo {author} {\bibfnamefont {D.}~\bibnamefont
  {Kim}}, \bibinfo {author} {\bibfnamefont {H.}~\bibnamefont {Lee}}, \bibinfo
  {author} {\bibfnamefont {B.}~\bibnamefont {Dabrowski}}, \bibinfo {author}
  {\bibfnamefont {S.}~\bibnamefont {Kolesnik}}, \bibinfo {author}
  {\bibfnamefont {J.}~\bibnamefont {Lee}}, \bibinfo {author} {\bibfnamefont
  {B.}~\bibnamefont {Kim}}, \bibinfo {author} {\bibfnamefont {B.}~\bibnamefont
  {Min}},\ and\ \bibinfo {author} {\bibfnamefont {J.-S.}\ \bibnamefont
  {Kang}},\ }\href {https://doi.org/https://doi.org/10.1103/PhysRevB.81.073101}
  {\bibfield  {journal} {\bibinfo  {journal} {Phys. Rev. B Condens. Matter}\
  }\textbf {\bibinfo {volume} {81}},\ \bibinfo {pages} {073101} (\bibinfo
  {year} {2010})}\BibitemShut {NoStop}%
\bibitem [{\citenamefont {Valenti}\ and\ \citenamefont {Park}(2026)}]{repo}%
  \BibitemOpen
  \bibfield  {author} {\bibinfo {author} {\bibfnamefont {A.}~\bibnamefont
  {Valenti}}\ and\ \bibinfo {author} {\bibfnamefont {I.}~\bibnamefont {Park}},\
  }\href@noop {} {\bibinfo {title} {{Trained model}}},\ \bibinfo {howpublished}
  {\url{https://github.com/agnes-valenti/mlDMFT}} (\bibinfo {year} {2026})\BibitemShut {NoStop}%
\end{thebibliography}

\end{document}